\documentclass{webofc}
\usepackage[caption = false]{subfig}
\usepackage[varg]{txfonts} 
\usepackage{wrapfig}
\usepackage{comment}

\newcommand{\pt}{\mbox{$p_T$}\xspace}

\newcommand{\pp}{\mbox{$p$$+$$p$}\xspace}

\newcommand{\jpsi}{\mbox{$J/\psi$}\xspace}
\newcommand{\psip}{\mbox{$\psi(2S)$}\xspace}
\newcommand{\auau}{\mbox{Au$+$Au}\xspace}

\newcommand{\pythia}{\mbox{\textsc{pythia8}}\xspace}

\begin{document}
\title{Heavy Flavor and Quarkonia Results \\from the PHENIX Experiment}

\author{\firstname{Krista} \lastname{Smith}\inst{1}\fnsep\thanks{\email{kristas@lanl.gov}} on behalf of the PHENIX Collaboration
}

\institute{Los Alamos National Laboratory}

\abstract{
The PHENIX experiment at RHIC has a unique large rapidity coverage ($1.2 < |y| < 2.2$) for heavy flavor studies in heavy-ion collisions. This kinematic region has a smaller particle density and may undergo different nuclear effects before and after the hard process when compared to mid-rapidity production. The latest PHENIX runs contain a large data set which allows, for the first time, the study of heavy flavor and $J/\psi$ flow at the large rapidity region in Au$+$Au collisions at $\sqrt{s_{_{NN}}}=$200 GeV. This measurement has the potential to reveal a medium evolution distinct from that known at mid-rapidity.}% This presentation will also report on the analysis status of non-prompt $J/\psi$  coming from B-meson decays at mid-rapidity in collisions.  This data can reach very low B-meson yields which is typically challenging to be described by perturbative quantum chromodynamic calculations.}

\maketitle

%==============================================================
\section{Introduction}
\label{sec:intro}
%==============================================================
Three recent heavy flavor and quarkonia analyses by the PHENIX Collaboration aim to investigate the following questions: 
 Do we see evidence for multi-parton interactions at RHIC energies?  Are there final state effects on charmonium production in \pp~collisions?  Is there evidence of mass ordering for charged hadron vs. open heavy flavor $v_2$ at forward rapidity?  Here we present the results from these ongoing studies and discuss the conclusions.

%==============================================================
\section{Data Set \& Experimental Setup}
\label{sec:data}
%==============================================================
The \pp data set used in two of the analyses was recorded in run year 2015 at a center of mass energy $\sqrt{s_{_{NN}}}$=200 GeV, while the \auau data set was recorded in 2014, also at the same center of mass energy per nucleon pair.  The integrated luminosity is 47 
pb$^{-1}$ for the 2015 \pp data set~\cite{PHENIX:2019gix} and 2.3 nb$^{-1}$ for the 2014 \auau data set~\cite{PHENIX:2022wim}.

All analyses have utilized the PHENIX Muon Arms, which includes the Forward Vertex Silicon Detector (FVTX), the Muon Tracker (MuTr), the Muon Arms magnet, and the Muon Identifier (MuID).  The FVTX consists of four layers of silicon strip stations for precision measurement of track trajectory and charged particle multiplicity.  The MuTr measures the momentum of charged tracks, and the MuID contains five layers of steel absorbers for hadron and muon separation.  At backward rapidity, the South Muon Arm covers -2.2 $<$ $\eta$ $<$ -1.2 in pseudorapidity while the forward arm, slightly larger, covers 1.2 $<$ $\eta$ $<$ 2.4 in pseudorapidity.

%==============================================================
\section{Multiplicity Dependent $J/\psi$ Production in \pp Collisions}
\label{sec:multiplicity}
\begin{figure}[h]
 % \begin{subfigure}
    \includegraphics[width=6cm]{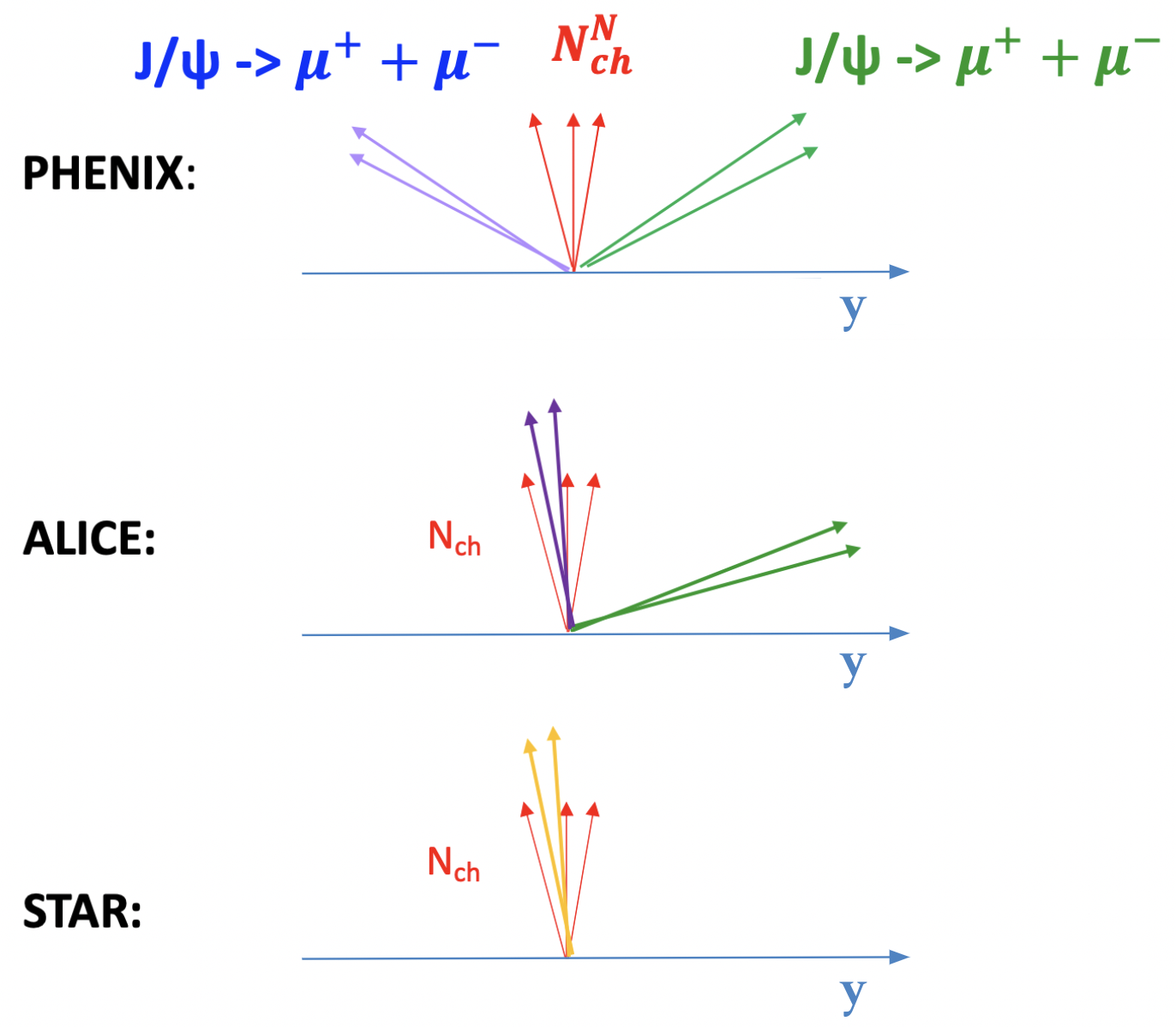}
%  \end{subfig}
  \hfill
%  \begin{subfig}
    \includegraphics[width=6cm]{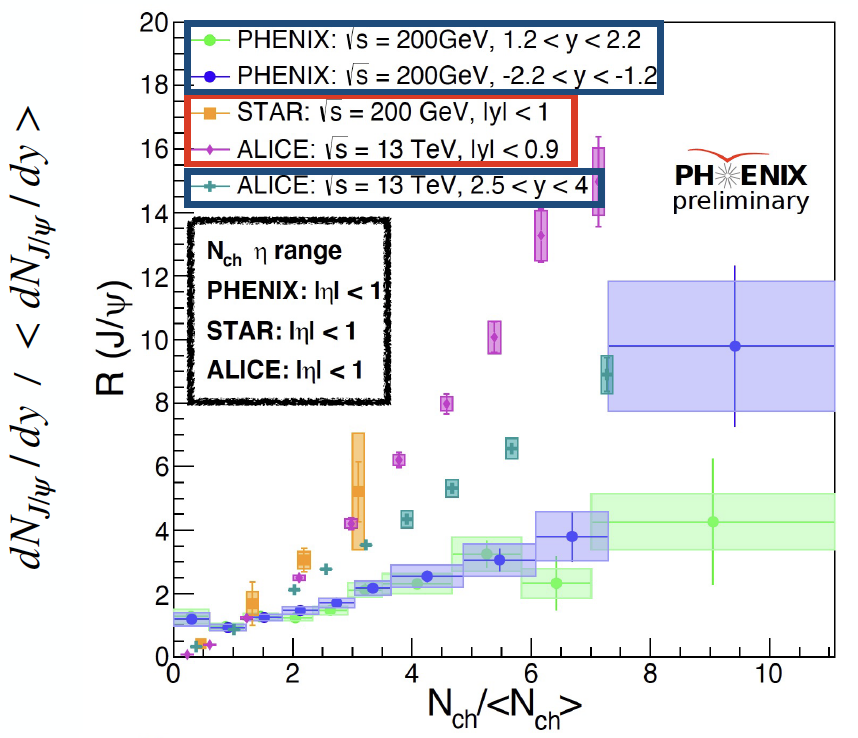}
    %\label{fig:f2}
 % \end{subfigure}
  \caption{Left:  Diagram showing the rapidity $y$ of the measured \jpsi compared to the measured charged particle tracks $N_{ch}$.  Right:  PHENIX \jpsi multiplicity measurements at forward (green data points) and backward (blue data points) are compared to ALICE~\cite{ALICE:2020msa} and STAR~\cite{STAR:2018smh} measurements at mid-rapidity in magenta and gold data points, and to ALICE measurements (teal data points) at forward rapidity.  Note that the ALICE and STAR charged particle tracks are measured at mid-rapidity.}
  \label{fig:figure1}
\end{figure}

The first PHENIX measurement of relative \jpsi~yields $R$ versus the normalized event charged particle multiplicity $N_{ch}$/$\langle N_{ch} \rangle$ in \pp~collisions at $\sqrt{s}$=200 GeV is shown in Figure~\ref{fig:figure1}.  The charged particle tracks have been measured in three different rapidity regions: forward, mid, and backward rapidity, as shown in the diagram on the left of Figure~\ref{fig:figure1}.   At RHIC energies, there are, on average, $\sim$5 charged particle tracks produced per Minimum Bias event; therefore, the \jpsi lepton pairs contribute to a large percentage of the total charged particle tracks.
To avoid biasing the charged particle multiplicity, the PHENIX Experiment has found that \jpsi tracks should be subtracted from (i.e., not included in) the charged particle multiplicity if the \jpsi is measured at the same rapidity~\cite{Shi:2023gnw}.  At LHC energies, however, the \jpsi lepton pairs do not constitute a large percentage of the overall charged particle tracks, and any bias from double counting (i.e., including the \jpsi~decay products in the multiplicity) is negligible.   

The PHENIX results (blue, green data points) are shown in the right plot after the \jpsi tracks have been subtracted from the charged particle multiplicity.  The dependence on multiplicity becomes less pronounced compared to the ALICE~\cite{ALICE:2020msa} and STAR~\cite{STAR:2018smh} results after this effect has been considered.  However, the multiplicity dependence does increase with increasing particle multiplicity, which is consistent with expectations of multi-parton interactions.

\begin{figure}[h]
\centering
\includegraphics[width=13cm,clip]{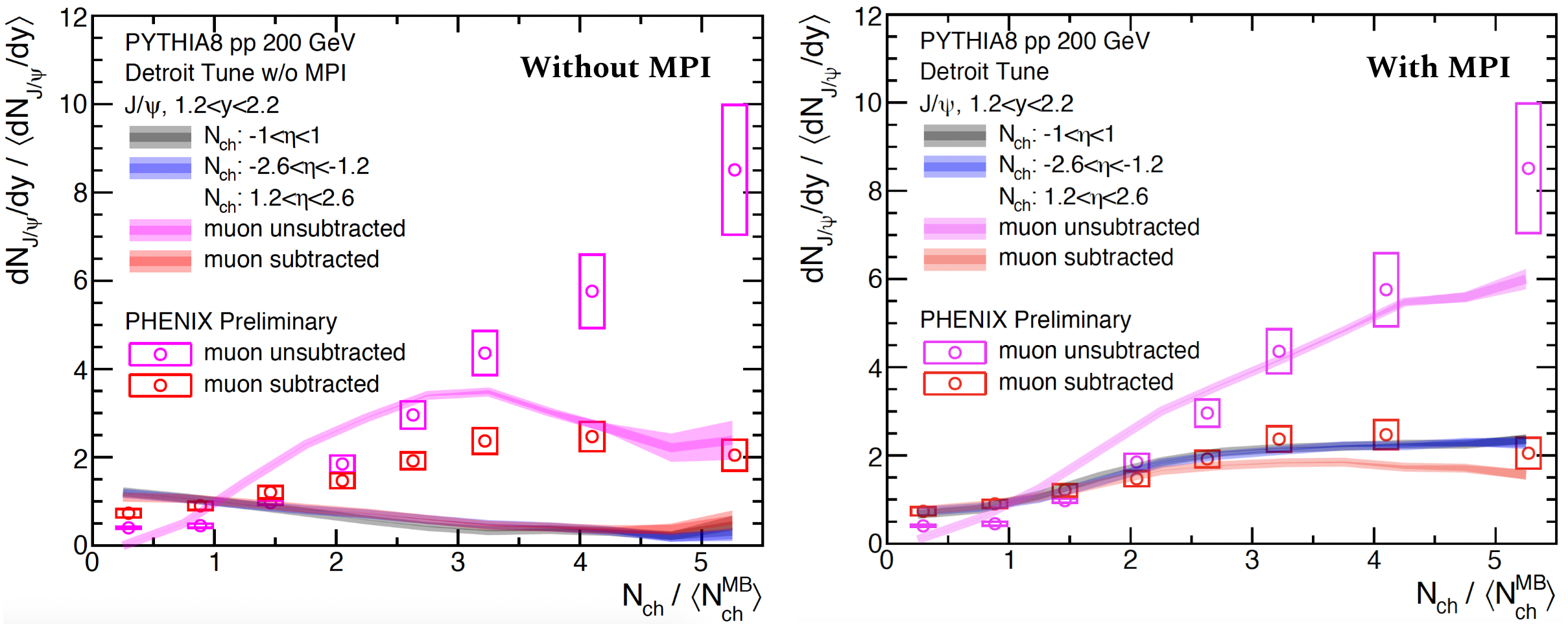}
\caption{The \jpsi multiplicity dependence is shown before (magenta data points) and after (red data points) the \jpsi~tracks are subtracted.  Both measurements are compared to \pythia Detroit tune~\cite{PYTHIA:81, Aguilar:2021sfa} calculations with (right) and without (left) multi-parton interactions turned on. }
\label{fig:shlim_detroit}      
\end{figure}

In Figure~\ref{fig:shlim_detroit}, the \jpsi multiplicity dependence is shown before (magenta data points) and after (red data points) the \jpsi~tracks have been subtracted.  Both measurements are compared to \pythia Detroit tune~\cite{PYTHIA:81, Aguilar:2021sfa} calculations both with (right) and without (left) multi-parton interactions turned on.  This study has concluded that the \jpsi multiplicity dependence is well described by the \pythia Detroit tune and that multi-parton interactions are required to reproduce PHENIX data.

%%%%%%%%%%%%%%%%%%%%%%%%% Figure 2
%%%%%%%%%%%%%%%%%%%%%%%%%

%%%%%%%%%%%%%%%%%%%%%%%%% Figure 3
%%%%%%%%%%%%%%%%%%%%%%%%%

%==============================================================
\section{$\psi(2S)$ to $J/\psi$ Ratio in \pp Collisions}
\label{sec:ratio}
\begin{figure}[h]
\centering
\includegraphics[width=13cm,clip]{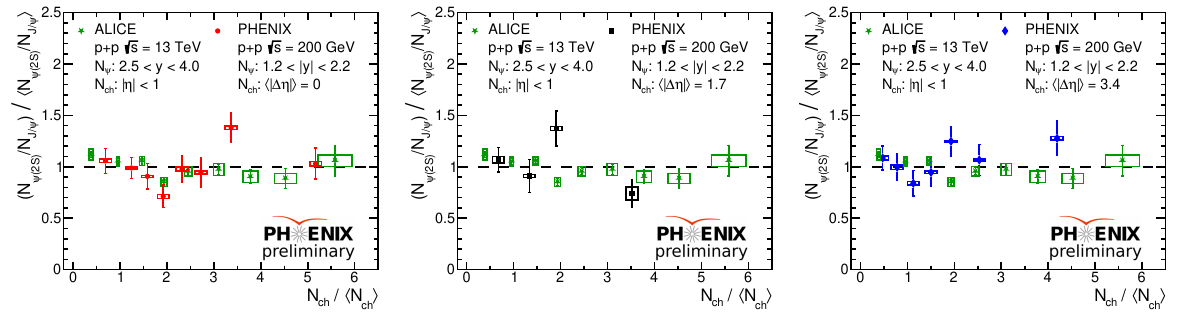}
\caption{The PHENIX \psip to \jpsi ratio in \pp collisions as a function of charged particle multiplicity (red, black, blue data points) is compared to ALICE measurements (green data points) at forward rapidity~\cite{ALICE:2022gpu}.  The PHENIX charged particle tracks are measured at forward rapidity (left), mid-rapidity (middle), and backward rapidity (right).  }
\label{fig:shlim_alice}      
\end{figure}

%==============================================================
Multiplicity-dependent studies in small system collisions can provide a testing ground for examining the onset of quark-gluon plasma-like effects.  Here the \psip to \jpsi ratio is measured in \pp collisions as a function of charged particle multiplicity at forward and backward rapidity.  Figure~\ref{fig:shlim_alice} shows the PHENIX measurements (red, black, blue data points) compared to ALICE measurements (green data points).  The left, middle, and right plots correspond to the PHENIX charged particle tracks measured at forward, mid, and backward rapidity, respectively.  Note that ALICE results have charged particle multiplicity measured at mid-rapidity only~\cite{ALICE:2022gpu}.  We find that the PHENIX ($\sqrt{s_{_{NN}}}$=200 GeV) and ALICE ($\sqrt{s_{_{NN}}}$=13 TeV) results are consistent, showing a weak multiplicity dependence more or less consistent with unity.  No evidence is found for \psip~final state effects in \pp~collisions at RHIC energies.

%%%%%%%%%%%%%%%%%%%%%%%%% Figure 4
%%%%%%%%%%%%%%%%%%%%%%%%%

%==============================================================
\section{Open Heavy Flavor $v_2$ in \auau Collisions}
\label{sec:heavy_flavor}
\begin{figure}[h]
\centering
\includegraphics[width=13cm,clip]{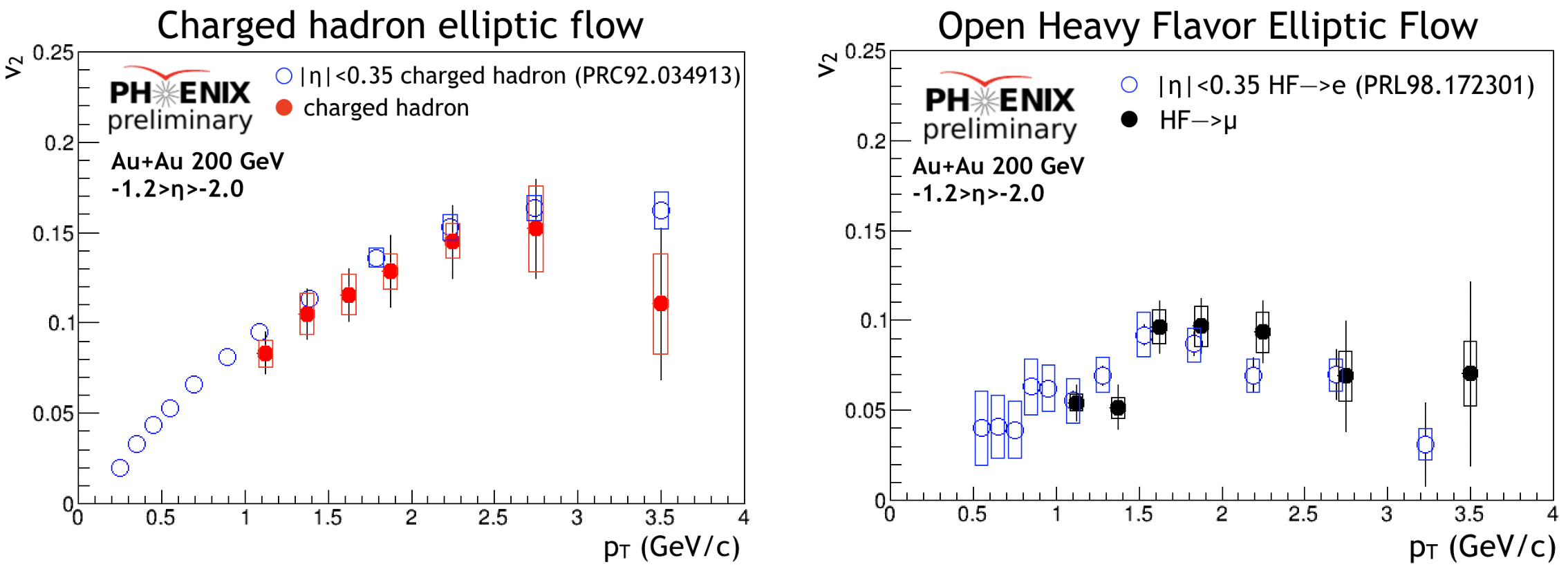}
\caption{Charged hadron elliptic flow (red data points, left) and open heavy flavor elliptic flow (black data points, right) in \auau collisions at forward rapidity as a function of \pt are compared to corresponding PHENIX elliptic flow measurements (blue data points) at mid-rapidity~\cite{PHENIX:2014yml, PHENIX:2006iih}.}
\label{fig:bran_v2}      
\end{figure}

%==============================================================
The rapidity dependence of elliptic flow measurements is of interest because particle production at mid versus forward rapidity could undergo different quark gluon plasma related effects, such as temperature and pressure gradients.  The first-ever RHIC measurement of open heavy flavor elliptic flow at forward rapidity is shown in Figure~\ref{fig:bran_v2}.  
The charged hadron elliptic flow (red data points, left) and open heavy flavor elliptic flow (black data points, right) in \auau collisions as a function of \pt are compared to corresponding PHENIX elliptic flow measurements (blue data points) at mid-rapidity~\cite{PHENIX:2014yml, PHENIX:2006iih}.  We find that the open heavy flavor $v_2$ is consistent with PHENIX mid-rapidity results. The light charged hadron $v_2$ (left) shows larger elliptic flow than the heavy-flavor $v_2$ over the entire \pt range of the data, and is also consistent with PHENIX mid-rapidity results.  From these measurements of charged hadron and open heavy flavor $v_2$, we conclude that similar quark gluon plasma effects are present at both mid and forward rapidities.

%%%%%%%%%%%%%%%%%%%%%%%%% Figure 4
%%%%%%%%%%%%%%%%%%%%%%%%%

%==============================================================
\section{Conclusion}
\label{sec:conclusion}
%==============================================================
%In conclusion, for \pp collision system results, we find that multiplicity dependent \jpsi~varies based on $\eta$ of the charged particle tracks.    
%In conclusion, for \pp collision system results, if the \jpsi~is measured at the same rapidity as the charged particle tracks, the \jpsi~decay products should be excluded from the multiplicity. 

In conclusion, for \pp collision system results, the \jpsi~multiplicity dependence increases with increasing particle multiplicity.  Additionally, PHENIX data is well described by the PYTHIA Detroit tune with Multi Parton Interactions turned on.  We conclude there is evidence for Multi Parton Interactions at RHIC energies.  The \psip~to \jpsi~ratio shows a weak dependence on multiplicity, and therefore no evidence is found for \psip~final state effects in \pp~collisions at RHIC.  In \auau collisions, the first RHIC measurement of open heavy flavor $v_2$ at forward rapidity is presented.  The results are consistent with PHENIX mid-rapidity measurements, suggesting similar quark gluon plasma effects (including temperature/pressure gradients) are present at both rapidities.  

%
% BibTeX or Biber users please use (the style is already called in the class, ensure that the "woc.bst" style is in your local directory)
% \bibliography{name or your bibliography database}

%==============================================================
%\begin{thebibliography}{}
%\bibliographystyle{elsarticle-num}
\bibliography{main.bib}
%==============================================================

\end{document}